\documentclass[12pt]{article}

\usepackage{sbc-template}

\usepackage{graphicx,url}

\usepackage{epstopdf}

\usepackage{amsmath}

\usepackage{subfigure}

\usepackage[latin1]{inputenc}

\newcommand*{\doi}[1]{\href{http://dx.doi.org/#1}{doi: #1}}

\title{Performance Assessment of WhatsApp and IMO on \\Android Operating System ($Lollipop$ and $KitKat$)\\ during VoIP calls using 3G or WiFi}

\author{R.C. de Oliveira\inst{1}, H.M. de Oliveira\inst{2}, 
        R.A. Ramalho and L.P.S. Viana\inst{3}} 

\address{
Computer Engineering -- Amazon State University (UEA)\\ 
Manaus, AM -- Brazil
\nextinstitute 
Statistics Department -- Federal University of Pernambuco (UFPE)\\ 
Recife, PE -- Brazil
\nextinstitute 
Escola Superior de Tecnologia, Amazon State University (UEA)\\ 
Manaus, AM -- Brazil\\
\email{rcoliveira@uea.edu.br, \{ronaldorar,lohana.anaiv\}@gmail.com}\\
\url{http://arxiv.org/a/deoliveira_h_1.html}}
\begin{document} 
\maketitle
\begin{abstract}
This paper assesses the performance of mobile messaging and VoIP connections. We investigate the CPU usage of \textit{WhatsApp} and \textit{IMO} under different scenarios. This analysis also enabled a comparison of the performance of these applications on two Android operating system (OS) versions: KitKat or Lollipop. Two models of smartphones were considered, viz. Galaxy Note 4 and Galaxy S4. The applications behavior was statistically investigated for both sending and receiving VoIP calls. Connections have been examined over 3G and WiFi. The handset model plays a decisive role in CPU usage of the application. $t$-tests showed that IMO has a better performance that WhatsApp whatever be the Android at a significance level 1\%, on Galaxy Note 4. In contrast, WhatsApp requires less CPU than IMO on Galaxy S4 whatever be the OS and access (3G/WiFi). Galaxy Note 4 using WiFi always outperformed S4 in terms of processing efficiency.

\end{abstract}

\section{Introduction}

Instant messaging and VoIP (voice over IP) for mobile phones are growing importance in the contemporary society. Instant messaging (IM) is a set of communication technologies used for text-based communication between two or more participants usually over the Internet \cite{Cherry}, \cite{Mattsson}. In particular, IM for mobile phones is becoming a worldwide fever \cite{OHara_et_al}, \cite{Butler}, \cite{Montag_et_al}. In performance evaluation of electronic devices is commonplace to build a base for comparison (\cite{Gosh_et_al}). Usually this database is constructed by applying tools that collect performance metrics (e.g. CPU, disk, memory and network statistics). Through such a baseline, the analyst can pinpoint where the drawbacks are, and carry out performance adjustments so as to improve the throughput of a given application. The choice of performance metrics, how performing the data collection, and data analysis are common steps of performance evaluation. We conducted a performance assessment of the WhatsApp as compared with the performance of IMO through 3G and Wifi, on different Android operating systems \cite{url}, \cite{Android}. The performance of such applications remains rather unexplored both from the theoretical viewpoint as well as in academia. See \cite{Church_Oliveira} for a comparison between WhatsApp and standard SMS.

\section{Materials and Methods}

The analysis delimited in this study is just VoIP on smartphones. The analysis carried out in this study is limited to monitoring the processing when instant messaging or voice call applications. The universe of study of this investigation is characterized by the scope of mobile devices operation. The field of study covered the transmission by wireless LAN (WiFi) or 3G networks \cite{Tanen_David}. It was not taken into account the coding, nor programming logic or source code of applications. The Android OS is a multitasking operating system for mobile devices, including smartphones and tablets, which have different versions \cite{Mehrotra_et_al}, \cite{Gilski_Stefa}. The main purpose is the analysis of cross-platform instant messaging for smartphones, viz. WhatsApp and IMO, with versions of Android, $KitKat$ (released Oct. 2013) and $Lollipop$ (released Nov.2014). See comparison between Android 5.0 Lollipop and Apple IOS 8 \cite{Raghuveer-Ananth}. For the present experiment we used an analysis tool, measurement techniques and statistical methods. The scope of the study was carefully designed to avoid interference from outside or assumptions that were not linked to the analysis. Moreover, for the proper ``background collecting'' of logs on mobile applications is essential to select software that is able to perform the data capture. Sampling tests were performed by selecting an appropriate tool to collect specific logs. Our choice fell on the $Little~Eye$ and thereafter it was possible to analyze the resources and ways processing \cite{little_eye}. Test devices were Samsung Galaxy S4 (S4) and Samsung Galaxy Note 4 (N4), both with heterogeneous hardware and which have been installed Android. To build the environment, it was also required to install and configure a wireless network as well as the availability of a 3G carrier chip. The tests involved the following steps: ($i$) Install the OS on the test device; ($ii$) Set up, install and operate software for testing; ($iii$) Set up, install and operate application software; ($iv$) Perform the collection of logs; ($v$) Handling the collected data; ($vi$) statistically analyze the data collected. ``Little Eye'' is a performance analysis and monitoring tool that can help to identify and fix bug in an application with Android versions from 2.3 \cite{little_eye}. It is a tool that supports metrics related to CPU, network resources, RAM, disk space, GPS and battery consumption. Its main features are:
\begin{itemize}
\item {\bf Measure:}  measuring the performance of applications gathering information about each feature used the device generating detailed statistics for each resource;
\item {\bf Analyze:} It brings information about the background of the data collected creating graphs and statistics for analysis;
\item {\bf Optimize:} Suggest improvements in resource consumption by optimizing the operating system.
\end{itemize}
\begin{figure}[ht]
\centering
\includegraphics[height=5.5cm, width=8.0cm]{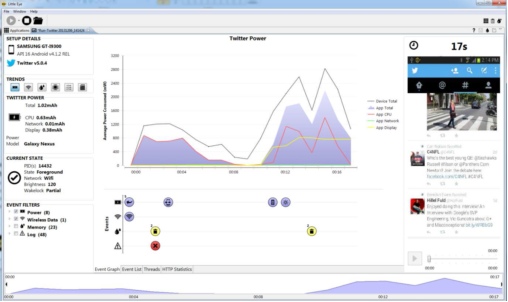}
\caption{ {\small Data Collection Interface in Little Eye (Source: $Little~Eye$).}}
\label{figura01}
\end{figure}
For tool installation the following requirements are necessary:\\
\indent{\textit{Java JRE or SDK - V 1.6 + (Java 6 or higher)} and \textit{Android SDK};}\\
\indent{\textit{USB debugging is enabled on the phone};}\\
\indent{\textit{Set the device to connect to as ``camera (PTP)'' rather than ``media device (MTP)''};}\\
\indent{\textit{Test device drivers are required when using Windows OS}.}\\
With everything set (hardware and software environments), Little Eye starts. Once started, it loads applications under test on the device. After listing all applications, simply select the application to be tested (WhatsApp or IMO), and then configure the measurements of interest, as illustrated in the following screen (Fig. \ref{figura01}).\\
The application under test is monitored with VoIP call duration of 1 minute, 5 minutes and 10 minutes. Data were collected during these periods. In a preliminary analysis, 30 calls with WhatsApp and IMO were refereed. The same test environment is applied to both Android $KitKat$ (KK) and $Lollipop$ (LL) systems, i.e., the same conditions are kept so there is no bias in results. Standard  hypothesis tests were conducted to ascertain a performance difference between IMO and WhatsApp applications. Two-tailed $t$-test for the population mean of IMO under a plenty of scenarios. Further convoluted tools are unneeded.
Let $\mu$ be the mean of CPU usage of the application during a 10 minutes VoIP call (mode sending or receiving). The statistical hypothesis at 1\% significance level ($\alpha=0.01$) were:
\begin{equation}
\label{hypothesis}
\begin{split}
null-hypothesis &~~~~H_0:  \mu_{IMO}=\mu_{Whatsapp}\\
alternative-hypothesis &~~~~H_1: \mu_{IMO} \neq \mu_{Whatsapp}.
\end{split}
\end{equation}
Also, left-tailed $t$-test showed evidence that IMO performance was higher than WhatsApp. $Bean~plot$  is also used to visualize performance data  \url{http://boxplot.tyerslab.com/}.
\begin{figure}[h]
\centering 
\subfigure[ref1][smartphone (Galaxy Note 4), OS (KitKat), transmission medium (3G)]{\includegraphics[width=7.0 cm]{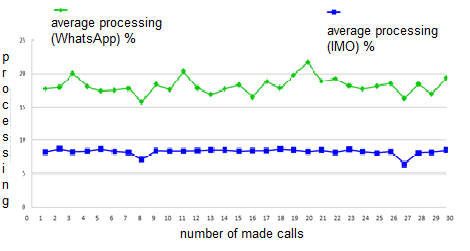}}
\qquad
\subfigure[ref2][smartphone (Galaxy Note 4), OS (KitKat), transmission medium (WiFi)]{\includegraphics[width=7.0 cm]{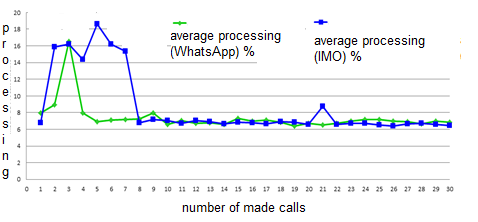}}
\caption{\small VoIP connection 10 minutes (WhatsApp / IMO), with smartphone transmitting messages.}
\label{figura02}
\end{figure}
\section{Performance of WHATSAPP and IMO}
Data collection was conducted using the Little Eye software as application performance analysis tool \cite{little_eye}. In this software, it is possible to collect smartphone application logs and analyze the processing consumption. This tool has a number of resources available to perform the analysis of applications, ranging from battery consumption to processing overhead. In these experiments, however, the scope has been narrowed to the study of behavior around the CPU. The test was carried out by collecting 30 calls lasting 10 minutes and the logs generated took the average for all sampling measures. These graphs show the applications are processed for use in receiving (also transmitting) a VoIP call through WhatsApp and the IMO using a WiFi connection as communication in both versions of Android \cite{url}. Figure \ref{figura02}, illustrates two instances of selected CPU requirements measures for achieving 30 calls, lasting 10 minutes. Each point is the average calculated from 30 samples. All correlations between performances in different scenarios were calculated: Higher performance correlations were obtained for the handset Galaxy S4 than for the Galaxy Note 4. The highest correlation coefficient among all tested scenarios was obtained for the Galaxy S4 with WhatsApp for transmitting/receiving text messages. Even requiring a memory load of roughly twice, the Wifi operation under the android LL had a similar behavior to a 3G transmission with the android version KK. For WhatsApp using the KitKat OS, the smartphone Galaxy S4 presented some correlation between 3G and Wifi. The KK android version yielded performance results not so sensitive to the selected network (3G or Wifi operation) and their memory requirements were pretty close. Still handling with WhatsApp on the device S4 (operating on WiFi), there is a performance correlation between the two android OS version, but the KK performance is roughly twice more efficient than the LL operation. Considering now the IMO application, in the 3G operation under Galaxy S4 smartphone, the general performance behavior is weakly sensitive to the version of the android system. Nevertheless, the performance of KK OS was approximately twofold more efficient than LL, as concerning 3G transmission. In contrast, the lowest correlation coefficient was found for KitKat in the two handset models, where the WhatsApp and IMO application performance for 3G calling were non correlated. Low correlations were also achieved for 3G connections on the smartphone Galaxy Note 4: the performance for KitKat and Lollipop were also uncorrelated.

\begin{figure}[ht]
\centering
\includegraphics[height=5.0cm, width=10.0 cm]{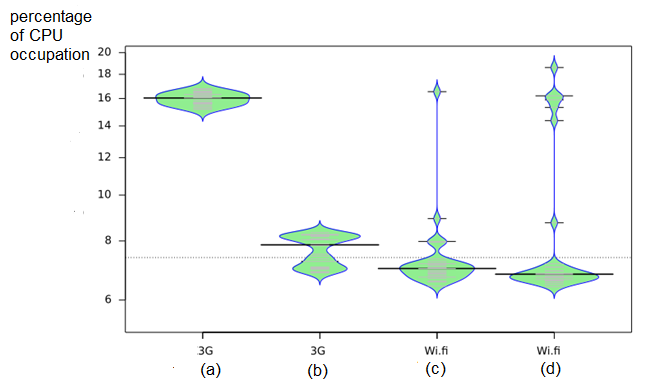}
\caption{ {\small Beanplot of CPU usage under the following scenario: Android OS $KitKat$, mobile device Samsung Galaxy Note 4 and transmitting calls mode. For 3G operating mode: a) Whatsapp and b) Imo. For WiFi operating mode: c) Whatsapp and d) IMO.}}
\label{figura03}
\end{figure}

\begin{table}[ht]
\centering
\caption{ {\small Average CPU usage for different scenarios. Smartphone Galaxy Models: N4 and S4. The calls were all made lasting 10 minutes. In each case, they were considered $N=30$ samples (each is an average obtained from 600 measurements). Values in parenthesis refer to the sample standard deviation. Significance level of $t$-test: $\alpha=0.01$}}.

\begin{center}
\begin{tabular}{| c| c c |c c | c c | c c |}
\hline
& 3G & KitKat & 3G & Lollipop &  WiFi & KitKat & WiFi & Lollipop\\
\hline
handset & WApp & IMO & WApp & IMO & WApp & IMO & WApp & IMO\\
\hline
Galaxy N4  & 15.97$^a$ & 7.66$^b$ & 15.64$^a$ & 6.51$^b$ & 7.40$^a$ & 8.68$^a$ & 16.18$^a$ & 8.42$^b$\\
           & (0.48)&(0.56)&(0.10)&(0.29)&(1.80)&(3.84)&(1.00)&(0.44)\\
\hline
Galaxy S4       &10.15$^a$&20.67$^b$&23.69$^a$&34.25$^b$&11.87$^a$&19.90$^b$&18.74$^a$&46.78$^b$\\
           & (1.64)&(0.52)& (0.68)&(3.07)&(1.11)&(0.57)&(7.80)&(1.19)\\
\hline
\end{tabular}\label{tabela01}
\end{center}
\end{table}
\noindent 
It was observed that a few specific moment, the processing occupation reached to zero. Sometimes this is expressed by display off (device screen hibernated,) it reduced the kernel processing consumption. It was noticed that there are three or more CPUs and the WhatsApp test has shown the using of a single CPU. This led to the idea that some features of the devices were being processed by other CPUs. In some cases it was perceived that the application crashes under Lollipop, but it is emphasized that troubles may have occurred during the logs collection. An example is the Internet itself, both 3G and WiFi, communication tool with the device or operating failures. In the beanplot (a variant of Tukey boxplot) shown in Fig. \ref{figura03}, one can see the behavior of CPU usage for measurements comparing the transmission medium (3G $\times$ WiFi) for WhatsApp and IMO. For 3G, a marked performance difference is observed between Whatsapp and IMO in the Samsung Note 4, showing a superior performance of IMO. In contrast, under WiFi, these differences are not remarkable. Table \ref{tabela01} (tx) and \ref{tabela02} (rx) present the statistics of average CPU requirements obtained in the pairwise measurements in order to compare the performance of WhatsApp and IMO. A marking with different letters (e.g. $a$ and $b$) indicates that the average CPU usage were different at a significance level of 1\% (so the hypothesis $H_0$ can be rejected). A pairwise comparison with the same letter ($a$ and $a$) indicates that the null hypothesis cannot be discarded at 1\%, i.e. there is no statistical evidence of performance difference between the two compared scenarios. In the first table, only the operating system version is changed (KitKat $\times$ Lollipop). In the second one, it is varied just the transmission medium (3G $\times$ WiFi). Null hypothesis (Eqn. (\ref{hypothesis})) is rejected at 1\% significance level in all cases, but \{KK,N4,Wifi\} where the performance of the IMO and WhatsApp is statistically indistinguishable ($t$=1.826, $p$-value=0.078). $p$-values were $p<10^{-5}$ in most cases. Also, left-tailed test have shown evidence to accept the hypothesis $\mu_{IMO} < \mu_{WhatsApp}$ (or $\mu_{WhatsApp}< \mu_{IMO}$). t-tests on Galaxy Note 4 have shown that IMO app (statistically) has a better performance that WhatsApp whatever the Android, at a significance level 1\%. In contrast, WhatsApp requires less CPU than IMO on Galaxy S4 at the same significance level, whatever the OS and the access network (3G/Wifi). Finally, Galaxy Note 4 using WiFi outperforms Galaxy S4 in terms of processing.
\begin{table}[ht]
\caption{ {\small Average CPU usage for different scenarios. Smartphone Galaxy Models: N4 and S4. Receiving message calls lasting 10 minutes. In each case, they were considered $N=30$ samples (each is an average obtained from 600 measurements). Values in parenthesis refer to the sample standard deviation. Significance level of $t$-test: $\alpha=0.01$}}
\begin{center}
\begin{tabular}{| c| c c |c c | c c | c c |}
\hline
& 3G & KitKat & 3G & Lollipop &  WiFi & KitKat & WiFi & Lollipop\\
\hline
handset & WApp & IMO & WApp & IMO & WApp & IMO & WApp & IMO\\
\hline
Galaxy N4  & 15.81$^a$ & 7.90$^b$ & 15.86$^a$ & 8.13$^b$ & 7.22$^a$ & 6.43$^a$ & 15.37$^a$ & 8.33$^b$\\
           & (1.20)    &(1.06)    & (0.10)    &(0.29)    &(0.47)    &(0.19)    &(1.04)     &(0.55)\\
\hline
Galaxy S4  & 11.57$^a$ & 19.75$^b$ & 23.41$^a$ & 34.92$^b$ & 10.61$^a$ & 19.89$^b$ & 22.17$^a$ & 31.85$^b$\\
           & (1.50)    & (0.25)    & (0.49)    & (2.68)    & (1.20)    & (0.30)    & (2.18)    & (5.76)\\
\hline
\end{tabular}\label{tabela02}
\end{center}
\end{table}

\section{Conclusions}

There is a marked increase in processing engendered by the version $Lollipop$ as compared to the $KitKat$. However, both on WiFi and 3G connections, there is insufficient data here to unveil the very reason, but the improvements made to the KitKat to create the Lollipop are focused on managing resources, such as energy consumption \cite{Saksonov}. Based on the results we can say that the operating system indirectly affects in the response in terms of CPU processing, although it may not be decisive. When comparing the same operating system on different chipsets we realize that the application directly contributes to the device performance. This claim comes from the realization that IMO on Galaxy S4 requires more CPU than WhatsApp, but on the other hand, this does not occur in the Galaxy Note 4. It is also observed that WiFi under Galaxy Note 4 has better performance than the Galaxy S4 in terms of processing, for both operating systems. It is quite likely to happen due to the CPU management, since each chip has its own managing way. In the 3G scenario, more CPU is required in both IMO and WhatsApp. It is assumed that the chipset combination, application development, Android OS and the network technology (WiFi/3G) is crucial in the CPU performance. The total processing using this application be given by the sum of CPU usage by the user (application) and CPU usage by the kernel generated by the application itself. Nevertheless, findings suggest the need for a more specific analysis from the perspective of resources exploited by each application. Ascertain the impact of energy consumption with the device update to the Android Lollipop version should also be carefully examined, since it is one of the notes issued for this release. It is so recommended as future research a deep investigation on energy consumption \cite{Murmuria_et_al}, \cite{Kundu_Kolin} achieved with the device to update the version Lollipop.  (\footnotesize{\url{http://cs.gmu.edu/~astavrou/research/Android_Power_Measurements_Analysis_SERE_12.pdf}}
\normalsize). Applications should have the chipset/OS as a key observance with a view on battery consumption.

\section*{Acknowledgment}
The authors thank to Samsung Ocean, Manaus, by valuable hardware support as well as to its general coordinator, Prof. Antenor Ferreira Filho for many interesting suggestions.

\end{document}